\documentclass{rnaastex}
\usepackage{amsmath}
\usepackage{url}

\begin{document}

\title{Noise-weighted angular differential imaging}

\correspondingauthor{Michael Bottom}
\email{mbottom@jpl.nasa.gov}

\author{Michael Bottom}
\affiliation{Jet Propulsion Laboratory, California Institute of Technology, Pasadena, CA, 91109 USA}

\author{Garreth Ruane}
\affiliation{Department of Astronomy, California Institute of Technology, MC 249-17, Pasadena, CA 91125, USA}

\author{Dimitri Mawet}
\affiliation{Jet Propulsion Laboratory, California Institute of Technology, Pasadena, CA, 91109 USA}
\affiliation{Department of Astronomy, California Institute of Technology, MC 249-17, Pasadena, CA 91125, USA}
\keywords{extrasolar planets --- high contrast imaging}


\section{Introduction} 
Angular differential imaging (ADI) \citep{2006ApJ...641..556M} is an observational technique in high contrast imaging where the telescope is used in pupil tracking mode so that the image of the sky rotates with respect to the optical surfaces.  Bright ``speckle'' light caused by optical errors remains fixed in the image, while planets and disks rotate with the sky.  The resulting dataset is then post-processed to remove the speckles, de-rotated to undo the sky motion, and median-collapsed to create a final data product.  The postprocessing algorithms to remove the speckles are an active area of research and beyond the scope of this note.  We consider the derotation and median-combination, where we show gains in signal-to-noise ratio are possible with a small change to the algorithm.  

\section{Against the median}
\begin{figure}[!htbp]
\epsscale{1}
\plotone{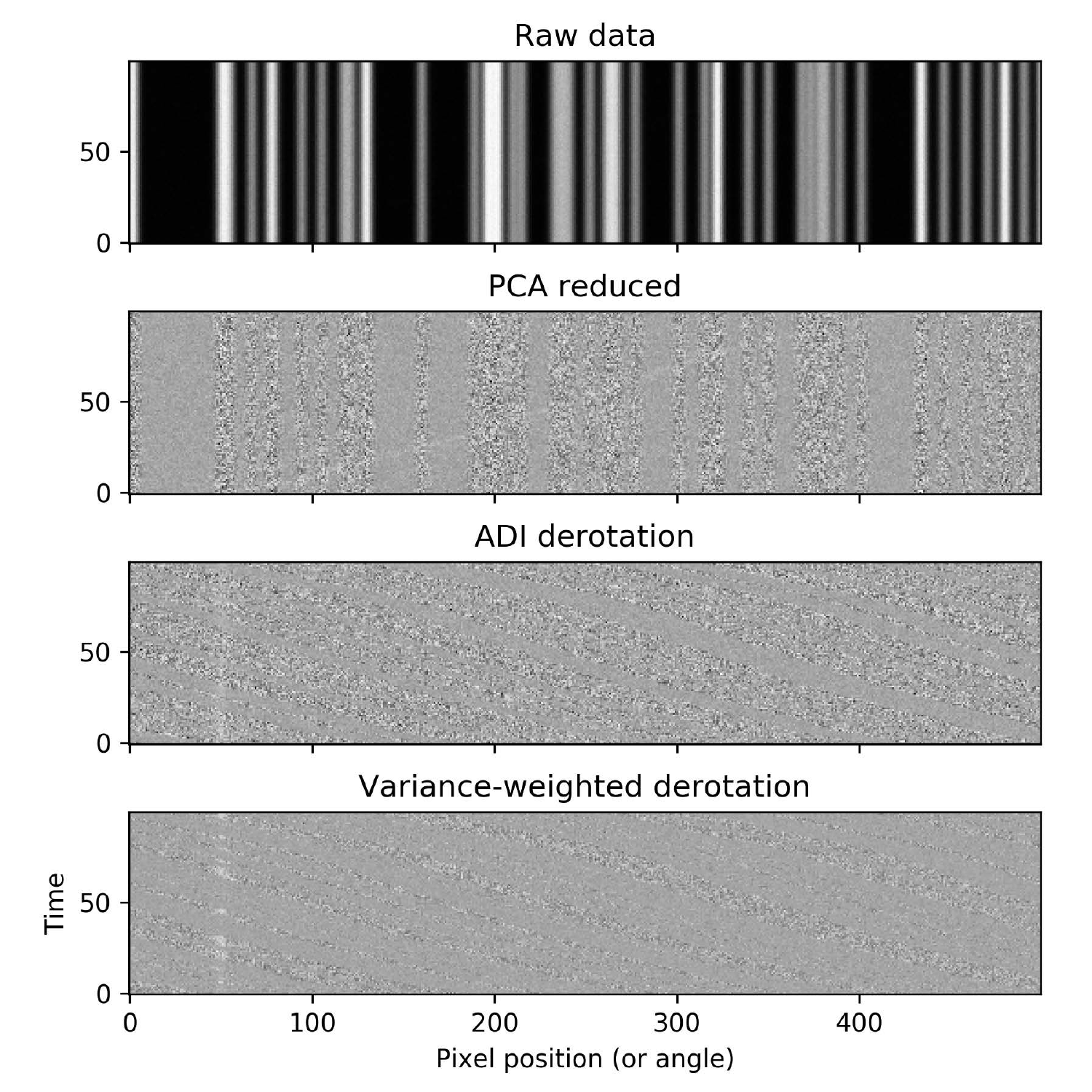}
\caption{1+1d reduction sequence of simulated data.  The axes are time ($y$) and pixel position ($x$).  (Top to bottom) ``cube'' with several bright, fixed speckles visible.  After a PCA reduction, the planet is visible ``rotating'' from left to right starting at pixel 50; note that while PCA (or any algorithm) can remove speckles, it cannot remove the Poisson variations from speckles.  With ADI derotation, the planet signal is aligned, but the residual noise is mixed together in the vertical direction; SNR is 7.  In the variance-weighted derotation, the planet is more clearly visible with noise accounted for and weighted; SNR is 9.5}
\label{reduction_1d}
\end{figure}

The main idea in this work is that the signal rotates through areas of the image with highly variable noise, and  derotating and median-collapsing the datacube ignores this, mixing the noise together (see Figure \ref{reduction_1d}).  We provide a simple algorithm to to locally weight the frames during combination to account for spatial noise variations. 

Consider two measurements of planet flux $F$, at just two different times (or positions, or rotation angles), which have different levels of noise.  We denote these as ($F_1, \sigma_1$) and ($F_2, \sigma_2$), and wish to combine these measurements to give a single flux.  Assume they are weighted by $w_1$ and $w_2$, so that the combined flux $F_{opt}$ is 

\begin{align}
w_1 F_1 + w_2 F_2 &= F_{opt}\\
w_1 + w_2 &= 1
\end{align}

\noindent The signal-to-noise is defined by $\mu(F_{opt})$/$\sigma(F_{opt})$.  This is given by

\begin{align}
SNR[F_{opt}]= \frac{w_1 F_1 + w_2 F_2 }{\sqrt{w_1^2 \sigma_1^2 + w_2^2 \sigma_2^2}}
\end{align}

\noindent By taking partials to maximize the SNR, one finds optimal weights of
\begin{align}
F_{opt} &= \frac{1}{\frac{1}{\sigma_1^2} + \frac{1}{\sigma_2^2}} \left( \frac{F_1}{\sigma_1^2} + \frac{F_2}{\sigma_2^2}\right)
\end{align}

\noindent This generalizes with multiple measurements to

\begin{align}
F_{opt} &= \frac{1}{ \sum_i \frac{1}{\sigma_i^2} }\sum_i \frac{F_i}{\sigma_i^2}
\label{theformula}
\end{align} 

This formula is well-known \citep{borenstein2011}, but rarely used with direct imaging datasets.  \citet{2017ApJ...842...14R} does use a noise weighting approach in a sophisticated matched-filter, but computes the noise in the neighborhood of the planet signal in the final reduced image.  This algorithm, presented for 2+1d data in the Appendix, is conceptually different as it operates on a pixel-by-pixel basis in the temporal axis of the datacube.

\section{Results with real data}
We tested this formula on real data using the VIP analysis pipeline \citep{2017AJ....154....7G} with the PCA, Annular PCA, and LLSG algorithms on data from Keck Infrared Vortex Coronagraph \citep{2017AJ....153...43S} and Palomar Stellar Double Coronagraph \citep{2016PASP..128g5003B}.  In all cases, the SNR improved, with 20-40\% gains being typical and 100\% gains possible.  We provide data and code to compare the median and variance-weighting for two such observations.\footnote{\url{https://mb2448.github.io/rnaas_data.html}}

With real data, ``dead'' detector pixels must be removed, as their low variance causes the denominator in Eq. \ref{theformula} to explode, making ``arcs'' in the derotated image.  Also, a ``running variance'' is useful in datasets with significant speckle evolution.

\section{Appendix}
The following steps implement Eq. \ref{theformula} in 1+2d reduced data with dimension $m\times n \times n$, where the $m$ refers to image number

\begin{enumerate}
\label{thelist}
\item Calculate the variance in the temporal ($m$) dimension of the reduced data $D$.
\item Create a variance cube $V$ that is $m\times n \times n$, with each frame a copy of 1).
\item Derotate the reduced data $D \rightarrow D_{rot}$
\item Derotate the variance cube $V \rightarrow V_{rot}$
\item Calculate $I_1 = \left[ \sum\limits_{m} (1/V_{rot}) \right]^{-1}$
\item Calculate $I_2 =\sum\limits_{m} (D_{rot}$/$V_{rot})$
\item The final image is $I_1 \circ I_2$, where $\circ$ is element-wise multiplication
\end{enumerate}

\section{Acknowledgements}
\acknowledgments
We thank the editor and two referees who reviewed a longer version of this note for \textit{Journal of Astronomical Telescopes, Instruments and Systems}.  One referee said it was a novel idea and should be published, the other said it was too minor to be published as a full journal article.  We agreed with both referees.

This work was partially performed at the Jet Propulsion Laboratory, California Institute of Technology, under contract with the National Aeronautics and Space Administration. (C) 2017.  All rights reserved.

\end{document}